\documentclass[superscriptaddress,reprint,pra]{revtex4-2}
\usepackage{amsmath,amssymb,amsthm}
\usepackage{newtxtext,newtxmath}
\usepackage{hyperref}
\usepackage{siunitx}
\usepackage{mathtools}
\usepackage{braket}
\usepackage{graphicx}
\usepackage{booktabs}
\usepackage{dcolumn}
\usepackage{bm}
\hypersetup{colorlinks=true,linkcolor=blue,citecolor=blue,urlcolor=blue}

\begin{document}

\title{Photon-Atom Granularity Noise Thermometry}
\author{Chen-Rong Liu}
\affiliation{College of Metrology Measurement and Instrument, China Jiliang University, Hangzhou 310018, China}

\author{Yixuan Wang}
\affiliation{College of Metrology Measurement and Instrument, China Jiliang University, Hangzhou 310018, China}

\author{Xiaowei Wang}
\affiliation{College of Metrology Measurement and Instrument, China Jiliang University, Hangzhou 310018, China}

\author{Chuang Li}
\affiliation{College of Metrology Measurement and Instrument, China Jiliang University, Hangzhou 310018, China}

\author{Mingti Zhou}
\affiliation{College of Metrology Measurement and Instrument, China Jiliang University, Hangzhou 310018, China}

\author{Runxia Tao}
\email{taorunxia@cjlu.edu.cn}
\affiliation{College of Metrology Measurement and Instrument, China Jiliang University, Hangzhou 310018, China}

\author{Hongwei Chen}
\email{hwchen9814@cjlu.edu.cn}
\affiliation{College of Metrology Measurement and Instrument, China Jiliang University, Hangzhou 310018, China}

\author{Ying Dong}
\email{yingdong@cjlu.edu.cn}
\affiliation{College of Metrology Measurement and Instrument, China Jiliang University, Hangzhou 310018, China}

\date{\today}

\begin{abstract}
We propose granularity noise thermometry (GNT), a fluctuation-based optical thermometry scheme that exploits the intrinsic fluctuations of susceptibility arising from atomic discreteness. The power spectral density of transmitted light exhibits an excess noise above the shot-noise limit that scales linearly with the photon-to-atom ratio $\mathcal{R}$. Consequently, varying the incident power (hence $\mathcal{R}$) yields the slope $\mathcal{K}$ of this linear scaling, which directly encodes the temperature. Closed-form expressions for the polarizability moments are derived via the plasma dispersion function, which yield distinct temperature scalings: $\mathcal{K}\propto P_{\mathrm{v}}(T)/T^2$ for thermal vapors and $\mathcal{K}\propto T^{2}$ for cold atoms. While practical implementation requires careful control of technical noise and system parameters, the present framework provides a noise-based pathway for optical thermometry using atomic ensembles.
\end{abstract}

\maketitle

\section{\label{sec:intro} Introduction}

Accurate temperature measurement plays a central role in fundamental physics and precision metrology, bridging the most stringent tests of physical law with the practical realization of primary standards. Over the years, a variety of primary thermometry schemes have been developed that realize thermodynamic temperature $T$ traceable to the International System of Units (SI)~\cite{mohr2016constants,fischer2018,stock2019si,pearce2026future}, each exploiting a distinct physical mechanism. Acoustic gas thermometry (AGT), for example, harnesses the adiabatic compressibility of an ideal gas, yielding the speed of sound $c_s \propto \sqrt{T}$. While achieving relative uncertainties on the order of $10^{-5}$ near room temperature, AGT typically requires large-volume resonators and corrections for non-ideal gas behavior~\cite{moldover1988measurement,moldover2014acoustic}. Johnson noise thermometry (JNT) represents a conceptually distinct paradigm. Based on Nyquist's theorem~\cite{nyquist1928thermal}, JNT measures the power spectral density (PSD) of electrical fluctuations across a resistor, which scales linearly with $k_{\mathrm{B}}T$ in the classical limit, thereby linking thermal fluctuations directly to $T$ by virtue of fundamental constants. State-of-the-art JNT achieves relative uncertainties on the order of $10^{-5}$, albeit with stringent requirements on ultra-low-noise electronics and precise impedance matching over broad bandwidths~\cite{white1996status,childs2000review,spietz2003primary,benz2011electronic,qu2019johnson,benz2024practical}. Concurrently, within optical metrology, Doppler broadening thermometry (DBT) retrieves $T$ from the Gaussian width of an atomic or molecular spectral line: the Doppler width $\Delta \nu_{\mathrm{D}}$ scales as $\sqrt{T}$. Near room temperature, DBT has demonstrated sub-10\,ppm uncertainty~\cite{zhang2022doppler,huang2025cavity}. However, it is not applicable at ultracold temperatures where $\Delta \nu_{\mathrm{D}} \to 0$. Moreover, in practice, DBT typically requires pressure extrapolation to eliminate collisional broadening.

\begin{figure}[t]
\centering
\includegraphics[width=\linewidth]{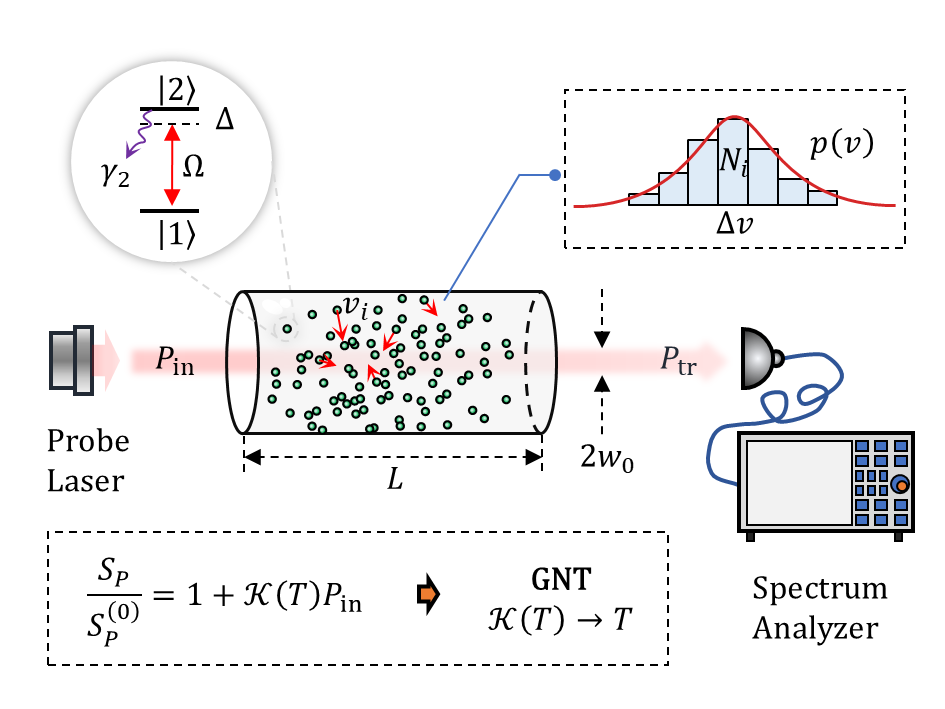}
\caption{Schematic of the GNT. A laser is amplitude-modulated (e.g., by an acousto-optic modulator) to vary the incident power $P_{\mathrm{in}}$ and thereby the photon flux. The beam with waist $w_0$ interacts with an atomic vapor cell of length $L$; for a cold atomic cloud, the physical cell is replaced by a trapped ensemble, and $L$ is understood as the effective optical length $L_{\mathrm{eff}}$ of the cloud. The transmitted power $P_{\mathrm{tr}}$ is measured by a low-noise photodetector, and its power spectral density is recorded by a spectrum analyzer. The measured PSD ratio follows the scaling law of Eq.~\eqref{eq:scaling}, where $\mathcal{R}$ denotes the photon‑to‑atom ratio (see text).}
\label{fig:gnt}
\end{figure}

Notably, JNT demonstrates that intrinsic fluctuations, traditionally regarded as noise, can serve as a direct thermometric resource. This concept naturally extends to light-matter interactions, where both photons and atoms are discrete. To see this, consider a probe laser illuminating an atomic ensemble. The beam has a finite volume (waist $w_0$, cell length $L$; see Fig.~\ref{fig:gnt}); the number of atoms within the beam is finite, and each atom's polarizability depends on its random thermal velocity, drawn from the Maxwell-Boltzmann distribution $p(v)$. The sample average over these atoms consequently exhibits sampling noise, which manifests as statistical fluctuations in the transmitted optical power. By analogy with JNT, one would expect the PSD of these fluctuations to encode the temperature. As yet, such a thermometer---one that extracts temperature from the fluctuation spectrum of light---has not been proposed, despite the recognition that atom-number fluctuations constitute a fundamental noise source in precision spectroscopy~\cite{dimarcq1994statistical,greiner2005probing}, specific theoretical proposals for noise thermometry in optical lattices~\cite{roscilde2014thermometry}, and recent experimental demonstrations of fluctuation-based thermometry via single-atom imaging~\cite{dixmerias2025fluctuation}.

In this work, we introduce granularity noise thermometry (GNT), which exploits the statistical fluctuations inherent to a finite atomic ensemble probed by a finite-volume optical beam. Building on a unified scaling law~\cite{liu2026granularity} that links the PSD of transmitted light to the photon‑to‑atom ratio, we demonstrate how it enables temperature retrieval applicable to both thermal vapors and cold atoms. Specifically, the transmitted PSD exceeds the shot‑noise limit by an amount that scales linearly with the photon-to-atom ratio $\mathcal{R}$. By varying the incident power (hence $\mathcal{R}$) and performing a linear fit, the slope $\mathcal{K}$ follows, expressed in terms of fundamental constants, known atomic parameters (transition dipole moment, decay rate), and the experimental geometry (beam waist $w_0$, cell length $L$). Incorporating the single-atom polarizability obtained from the optical Bloch equations, we can evaluate the slope $\mathcal{K}$ analytically using the plasma dispersion function~\cite{fried1961plasma}. For thermal vapors, where the Doppler width far exceeds the homogeneous linewidth, $\mathcal{K} \propto P_{\mathrm{v}}(T)/T^2$ emerges; for a fixed-number cold atomic cloud, where the homogeneous linewidth dominates, $\mathcal{K} \propto T^2$. The statistical precision of $\mathcal{K}$ and the associated temperature uncertainty are evaluated by standard error propagation~\cite{degen2017quantum,liu2026fisher}, complemented by a Fisher information analysis. In practice, the experimental setup is straightforward and compact (see Fig.~\ref{fig:gnt}).

The remainder of this article is organized as follows. Section~\ref{sec:stat} develops the statistical framework and derives the fundamental scaling relation. Section~\ref{sec:twolevel} presents the two-level atom model and establishes the key parameter $b_0$ that governs the temperature dependence. Section~\ref{sec:temp} derives explicit temperature scalings for thermal vapors and cold atoms, including the asymptotic $b_0\ll1$ and $b_0\gg1$ regimes. Section~\ref{sec:discussion} examines the achievable precision, experimental considerations, and limitations of the approach. Concluding remarks are given in Sec.~\ref{sec:conclusion}. Appendices contain detailed derivations.

\section{\label{sec:stat} Statistical framework for atomic ensembles}

The foundation of our thermometry scheme lies in understanding the statistical properties of the optical response of the atoms contained within the probe beam. Unlike conventional treatments that consider a continuous medium, we explicitly account for the finite number of atoms contributing to the signal and the random distribution of their kinematic states. This granularity is the very source of the noise we aim to exploit.

\subsection{Susceptibility as a sample statistic}

Consider an ensemble of $N$ atoms contained within a probe beam of volume $V_{\mathrm{bm}}$ (see Fig.~\ref{fig:gnt}). For a thermal vapor, $V_{\mathrm{bm}} = \pi w_0^2 L$ is the volume of the beam within the cell; for a cold atomic cloud probed by a beam that fully encompasses the cloud, $V_{\mathrm{bm}}$ is the volume of the cloud itself. Each atom possesses a polarizability $\alpha(v)$ that depends on a state variable $v$ (e.g., velocity). The sample-averaged polarizability is thus given by
\begin{equation}
\bar{\alpha}_N = \frac{1}{N}\sum_{i=1}^N \alpha(v_i),
\end{equation}
where the $v_i$ are independent draws from the distribution $p(v)$. Assuming the atomic ensemble is spatially uniform and isotropic within the probe volume, linear response theory~\cite{jackson1999classical} gives the susceptibility as
\begin{equation}\label{eq:barchi}
\bar{\chi} = \frac{n}{\epsilon_0}\,\bar{\alpha}_N,
\end{equation}
where $n$ is the atomic density and $\epsilon_0$ the vacuum permittivity. For a spatially inhomogeneous cloud, $n$ should be understood as the average density over the probe volume, and Eq.~\eqref{eq:barchi} remains valid provided the probe intensity is approximately uniform over the cloud.

For an atomic vapor, the number of atoms in the probe volume $V_{\mathrm{bm}}$ follows a Poisson distribution with mean $\bar{N}_{\mathrm{at}} = nV_{\mathrm{bm}}$. The statistical properties of $\bar{\chi}$ follow from the law of total expectation and the law of total variance. Let $\bm{\Sigma}_{\alpha}$ denote the $2\times2$ covariance matrix of the vector $(\alpha_R,\alpha_I)^{\mathsf{T}}$. For the sample mean $\bar{\alpha}_N$, we obtain
\begin{equation}
\bm{\Sigma}_{\bar{\alpha}} = \bm{\Sigma}_{\alpha} \cdot \mathbb{E}\!\left[\frac{1}{N}\right].
\end{equation}
For $\bar{N}_{\mathrm{at}}\gg1$, which is typical in experiments, we thus have $\mathbb{E}[1/N] \approx 1/\bar{N}_{\mathrm{at}}$. Consequently, the diagonal entries of $\bm{\Sigma}_{\bar{\alpha}}$ give the variance for each quadrature $q\in\{R,I\}$:
\begin{equation}\label{eq:var_chi}
\operatorname{var}(\bar{\chi}_q) \approx \frac{1}{\bar{N}_{\mathrm{at}}}\underbrace{\frac{n^2}{\epsilon_0^2}\operatorname{var}_v[\alpha_q]}_{\mathcal{V}_q},
\end{equation}
where $\mathcal{V}_q$ isolates the intrinsic atomic fluctuations, and $\operatorname{var}_v[\cdot]$ denotes the variance with respect to $p(v)$. Equation~\eqref{eq:var_chi} shows that the fluctuations of the susceptibility are inversely proportional to the average number of atoms, a hallmark of a statistical sampling process. For a cold atomic sample with a fixed number of atoms $N_{\mathrm{at}}$, the situation is even simpler: the density may be treated as constant during the measurement~\cite{metcalf1999laser,wolswijk2025trapping}. In this case, $N$ is no longer a random variable, and the variance in Eq.~\eqref{eq:var_chi} holds exactly with $\bar{N}_{\mathrm{at}}$ replaced by $N_{\mathrm{at}}$.

\subsection{Unified scaling law}

The Beer-Lambert law governs the transmitted power through the atomic ensemble, encoding the susceptibility fluctuations in an experimentally accessible observable. In the case of a thermal vapor contained in a cell of length $L$, this takes the standard form
\begin{equation}
P_{\mathrm{tr}} = P_{\mathrm{in}} \exp\!\left(-\frac{2\pi L}{\lambda}\bar{\chi}_I\right),
\end{equation}
where $\bar{\chi}_I$ is the imaginary part of the sample-averaged susceptibility [Eq.~\eqref{eq:barchi}], and $P_{\mathrm{tr}}$ and $P_{\mathrm{in}}$ denote, respectively, the transmitted and incident optical powers (Fig.~\ref{fig:gnt}). As detailed in Appendix~\ref{app:psd}, the statistics of the transmitted power over an integration time $\Delta t$ determine the measured power spectral density. For a cold atomic cloud, there is no confining cell, yet the same exponential form remains valid provided $L$ is replaced by an effective optical length
\begin{equation}\label{eq:Leff}
L_{\mathrm{eff}} = \frac{N_{\mathrm{at}}}{\bar{n} \pi w_0^2} = \frac{V_{\mathrm{cl}}}{\pi w_0^2},
\end{equation}
where $\bar{n} = N_{\mathrm{at}}/V_{\mathrm{cl}}$ is the average density over the probe volume, with $V_{\mathrm{cl}}$ the cloud volume and $w_0$ the beam waist~\cite{metcalf1999laser,wolswijk2025trapping}. This identification follows from equating the optical depth of the non-uniform cloud, $\int{\sigma_0 n(\mathbf{r})\,dz}$ (with $\sigma_0$ the resonant absorption cross section), to the homogeneous-medium expression $\sigma_0 \bar{n} L_{\mathrm{eff}}$, ensuring consistency of the transmitted power fluctuations in both configurations.

To characterize the noise in the transmitted power, we consider a measurement time $\Delta t$. The number of detected photons $\xi$ follows a Poisson distribution with mean $\bar{N}_{\mathrm{ph}} = P_{\mathrm{in}}\Delta t/(\hbar\omega)$, as is appropriate for a coherent-state probe field, and the susceptibility $\bar{\chi}_I$ has variance given by Eq.~\eqref{eq:var_chi}. These two random variables are statistically independent. The variance of the transmitted power, and hence the power spectral density $S_P = 2\Delta t \operatorname{var}[P]$~\cite{oppenheim1999discrete,mandel1996optical}, then yields a remarkably simple scaling law for the normalized PSD (see Appendix~\ref{app:psd}):
\begin{equation}\label{eq:scaling}
\frac{S_P}{S_P^{(0)}} = 1 + \mathcal{R} \mathcal{I},
\end{equation}
where $S_P^{(0)} = 2\hbar\omega P_{\mathrm{tr}}^2/P_{\mathrm{in}}$ is the shot-noise level in the absence of atomic fluctuations ($\sigma_I^2 \to 0$), which is determined by the incident and transmitted optical powers as derived in Appendix~\ref{app:psd}. The photon-to-atom ratio $\mathcal{R}$ and the intrinsic fluctuation parameter $\mathcal{I}$ are given by
\begin{align}
\mathcal{R} &= \frac{\bar{N}_{\mathrm{ph}}}{N_{\mathrm{at}}}, \\[4pt]
\mathcal{I} &= \left(\frac{2\pi L}{\lambda}\right)^{\!2}\frac{n^2}{\epsilon_0^2}\operatorname{var}_v[\alpha_I].
\end{align}
Here $N_{\mathrm{at}}$ denotes the average atom number $\bar{N}_{\mathrm{at}}$ for a vapor, or the fixed atom number $N_{\mathrm{at}}$ for a cold cloud; $L$ is the cell length or the effective optical length $L_{\mathrm{eff}}$, as appropriate; $n$ is the uniform density or the average density $\bar{n}$ over the probe volume. The parameter $\mathcal{I}$ encapsulates all temperature dependence through the velocity distribution and atomic density. 

In the following sections, we exploit the linear dependence on $\mathcal{R}$ to isolate the intrinsic fluctuation parameter $\mathcal{I}$ by varying the incident power and fitting the resulting linear scaling, thereby accessing the temperature information encoded in the atomic ensemble.

\section{\label{sec:twolevel} Two-level atom polarizability}

To make quantitative predictions from the scaling law, we require a model for the single-atom polarizability, whose variance determines the intrinsic fluctuation parameter $\mathcal{I} \propto \operatorname{var}_v[\alpha_I]$ in Eq.~\eqref{eq:scaling}. We consider a two-level atom with ground state $|1\rangle$ and excited state $|2\rangle$, illuminated by a probe laser (see Fig.~\ref{fig:gnt}) of frequency $\omega$ and Rabi frequency $\Omega$. In the rotating frame, the Hamiltonian including the Doppler shift reads
\begin{equation}
H_R = \frac{\hbar}{2}\begin{pmatrix}
0 & \Omega \\
\Omega & -2(\Delta - kv)
\end{pmatrix},
\end{equation}
where $\Delta = \omega - \omega_{21}$ is the detuning from resonance, $k = 2\pi/\lambda$ is the probe wave number, and $v$ the atomic velocity along the beam direction.

The atom undergoes spontaneous decay from $|2\rangle$ to $|1\rangle$ at rate $\gamma_2$,
transit-time broadening~\cite{sagle1996measurement} at rate $\Gamma_t = \bar{v}/w_0$
(with $\bar{v} = \sqrt{8/\pi}\,v_{\mathrm{th}}$ the mean thermal speed), and
collision-induced dephasing~\cite{jabbour1995measurement,pitz2010pressure,steck2025Cesium} at rate
$\Gamma_c = n\sigma_{\mathrm{coll}}\bar{v}_{\mathrm{rel}}$
(with $\bar{v}_{\mathrm{rel}} = 4v_{\mathrm{th}}/\sqrt{\pi}$ the mean relative speed).
Here, the characteristic thermal velocity is $v_{\mathrm{th}} = \sqrt{k_B T/m}$. The resulting relaxation is captured by a master equation for $\rho$ with the corresponding Lindblad terms; the full master equation and its steady-state solution via the Drazin inverse technique~\cite{miller2024rydiqule,nagib2025fast} are detailed in Appendix~\ref{app:master}. The two dephasing rates combine additively into the homogeneous linewidth $\Gamma = \Gamma_t + \Gamma_c$. The steady-state coherence $\rho_{21}(v)$ then follows analytically and takes the compact rational form
\begin{equation}\label{eq:rho21v}
\rho_{21}(v)=\frac{\gamma_{2}\left[-i(\gamma_{2}+2\Gamma) + 2(\Delta - kv)\right]\Omega}
                  {\gamma_{2}\left[(\gamma_{2}+2\Gamma)^{2} + 4(\Delta - kv)^{2}\right] + 2(\gamma_{2}+2\Gamma)\Omega^{2}}.
\end{equation}
The polarizability $\alpha(v)$ is proportional to $\rho_{21}(v)$ via linear response
theory~\cite{steckQuantumAtomOptics} as
$\alpha(v) = (2|\bm{\epsilon}\cdot\bm{d}_{21}|^{2}/\hbar\Omega)\,\rho_{21}(v)$
(see Appendix~\ref{app:master}), where $\bm{\epsilon}$ is the probe polarization
vector and $\bm{d}_{21}$ the dipole matrix element. Introducing the dimensionless velocity $\xi = v/v_{\mathrm{th}}$, a partial-fraction
decomposition of Eq.~\eqref{eq:rho21v} reduces the polarizability to a sum of simple poles:
\begin{equation}\label{eq:alpha}
\alpha(\xi) = \frac{c_1}{\xi - r_1} + \frac{c_2}{\xi - r_2},
\end{equation}
with the poles given by
\begin{equation}\label{eq:r12}
r_{1,2}= \frac{\Delta}{k v_{\mathrm{th}}} \pm i\,\frac{\gamma_2+2\Gamma}{2k v_{\mathrm{th}}}\,\sqrt{1+x},
\end{equation}
and the real coefficients
\begin{equation}\label{eq:c12}
c_{1,2}=C_0\left(1\pm \frac1{\sqrt{1+x}}\right), \quad C_0 = -\frac{|\bm{\epsilon}\cdot\bm{d}_{21}|^{2}}{2\hbar k v_{\mathrm{th}}}.
\end{equation}
Here, the saturation parameter $x$ is
\begin{equation}\label{eq:x}
x = \frac{2\Omega^2}{\gamma_2(\gamma_2+2\Gamma)} = \frac{I}{I_{\mathrm{sat}}}.
\end{equation}
For $x \ll 1$, the atom remains in the unsaturated regime, which defines the weak-probe limit adopted throughout this work. The pole structure Eq.~\eqref{eq:alpha} permits analytical integration over the Maxwell-Boltzmann distribution via the plasma dispersion function~\cite{fried1961plasma}; the statistical moments of the polarizability can thus be obtained analytically, as collected in Appendix~\ref{subapp:moments}, and from them the required variance $\operatorname{var}_v[\alpha_I]$ follows directly.

For the remainder of this work, we focus on the experimentally simplest and most relevant case: resonant excitation to maximize the signal, and the weak-probe limit to avoid saturation nonlinearities. Under resonant excitation ($\Delta=0$) and in the weak-probe limit ($x\ll1$), the poles simplify to purely imaginary values $r_{1,2} = \pm i b_0$, and the coefficients reduce to $c_1 = 2C_0$, $c_2 = 0$, where
\begin{equation}\label{eq:b0}
b_0 \equiv \frac{\gamma_2+2\Gamma}{2k v_{\mathrm{th}}}.
\end{equation}
The parameter $b_0$, i.e., the ratio of the homogeneous linewidth to the Doppler width, fully governs the variance $\operatorname{var}_v[\alpha_I]$ and thus the temperature
dependence of the GNT signal. Substituting these simplified forms into the general moment expressions of Appendix~\ref{subapp:moments} yields the exact variance
\begin{equation}\label{eq:var_res}
\operatorname{var}_v[\alpha_I] = 2C_0^{2}\!\left[
\frac{\Im Z_0}{\sqrt{2}}\left(\frac1{b_0}-b_0\right)+1-(\Im Z_0)^{2}\right],
\end{equation}
with $Z_0 = Z(i b_0/\sqrt{2})$ and
\begin{equation}
\Im Z_0 = \sqrt{\pi}\,e^{b_0^{2}/2}\operatorname{erfc}\!\left(\frac{b_0}{\sqrt{2}}\right).
\end{equation}
The temperature dependence of $\operatorname{var}_v[\alpha_I]$ is thus entirely contained
in $b_0$. Its asymptotic analysis, presented in Appendix~\ref{app:asymp}, provides the
explicit scalings used throughout Sec.~\ref{sec:temp}. Unless stated otherwise, we adopt
these conditions throughout the remainder of this work.

\section{\label{sec:temp} From scaling law to temperature extraction}

The scaling law Eq.~\eqref{eq:scaling} expresses the excess noise in terms of the photon-to-atom ratio $\mathcal{R}$ and the intrinsic fluctuation parameter $\mathcal{I}$. Linking these quantities to experimentally controllable parameters depends on the physical realization of the atomic ensemble. We treat the two paradigmatic cases separately.

\subsection{Slope determination}

\subsubsection{Atomic vapor}

In an atomic vapor, the effective interaction time per atom is the transit time
\begin{equation}
\tau_{\mathrm{tr}} = \frac{2w_0}{\bar{v}} = \frac{2}{\Gamma_t},
\end{equation}
where $\Gamma_t$ is defined in Sec.~\ref{sec:twolevel}. From kinetic theory, the atomic flux through the transverse cross section of the probe beam is
\begin{equation}
\Phi_{\mathrm{at}} = \frac14 n\bar{v}A,
\end{equation}
with $A = 2\pi w_0L$ the lateral area, and the corresponding probe volume is $V_{\mathrm{bm}} = \pi w_0^2 L$. Here, we approximate the beam as a cylinder, which is an excellent approximation when the Rayleigh range exceeds the vapor cell length $L$. The average number of atoms in the probe volume follows as
\[
\bar{N}_{\mathrm{at}} = \Phi_{\mathrm{at}} \tau_{\mathrm{tr}} = n\pi w_0^2 L,
\]
which coincides with $\bar{N}_{\mathrm{at}} = nV_{\mathrm{bm}}$, reflecting the dynamical equilibrium of atoms entering and leaving the probe volume.

Meanwhile, the photon flux is simply $\Phi_{\mathrm{ph}} = P_{\mathrm{in}}/(\hbar\omega)$. For integration time $\Delta t$, the expectation number of detected photons is $\bar{N}_{\mathrm{ph}} = \Phi_{\mathrm{ph}}\Delta t$. The photon-to-atom ratio then becomes
\begin{equation}
\mathcal{R} = \frac{\bar{N}_{\mathrm{ph}}}{\bar{N}_{\mathrm{at}}} = \frac{\Phi_{\mathrm{ph}}}{\Phi_{\mathrm{at}}} \cdot \frac{\Delta t}{\tau_{\mathrm{tr}}}.
\end{equation}
A natural choice is to set the integration time equal to the transit time, $\Delta t = \tau_{\mathrm{tr}}$, so that each atom interacts with the probe beam for exactly one measurement window. In this case, $\mathcal{R}$ simplifies to the flux ratio $\mathcal{R} = \Phi_{\mathrm{ph}}/\Phi_{\mathrm{at}}$. Substituting the explicit fluxes into Eq.~\eqref{eq:scaling} yields
\begin{equation}\label{eq:scaling2K}
\frac{S_P}{S_P^{(0)}} = 1 + \frac{4\mathcal{I}}{\hbar\omega n\bar{v}A}\,P_{\mathrm{in}} \equiv 1 + \mathcal{K}_{\mathrm{v}} P_{\mathrm{in}},
\end{equation}
where 
\begin{equation}\label{eq:Kv}
\mathcal{K}_{\mathrm{v}} = \frac{\sqrt{2\pi}L}{\epsilon_0^{2}\hbar c w_0\lambda}\,\frac{n \operatorname{var}_v[\alpha_I]}{v_{\mathrm{th}}}
	\end{equation}
is the slope that can be extracted from a linear fit of the measured PSD ratio versus $P_{\mathrm{in}}$.

\begin{figure}[htbp]
\centering
\includegraphics{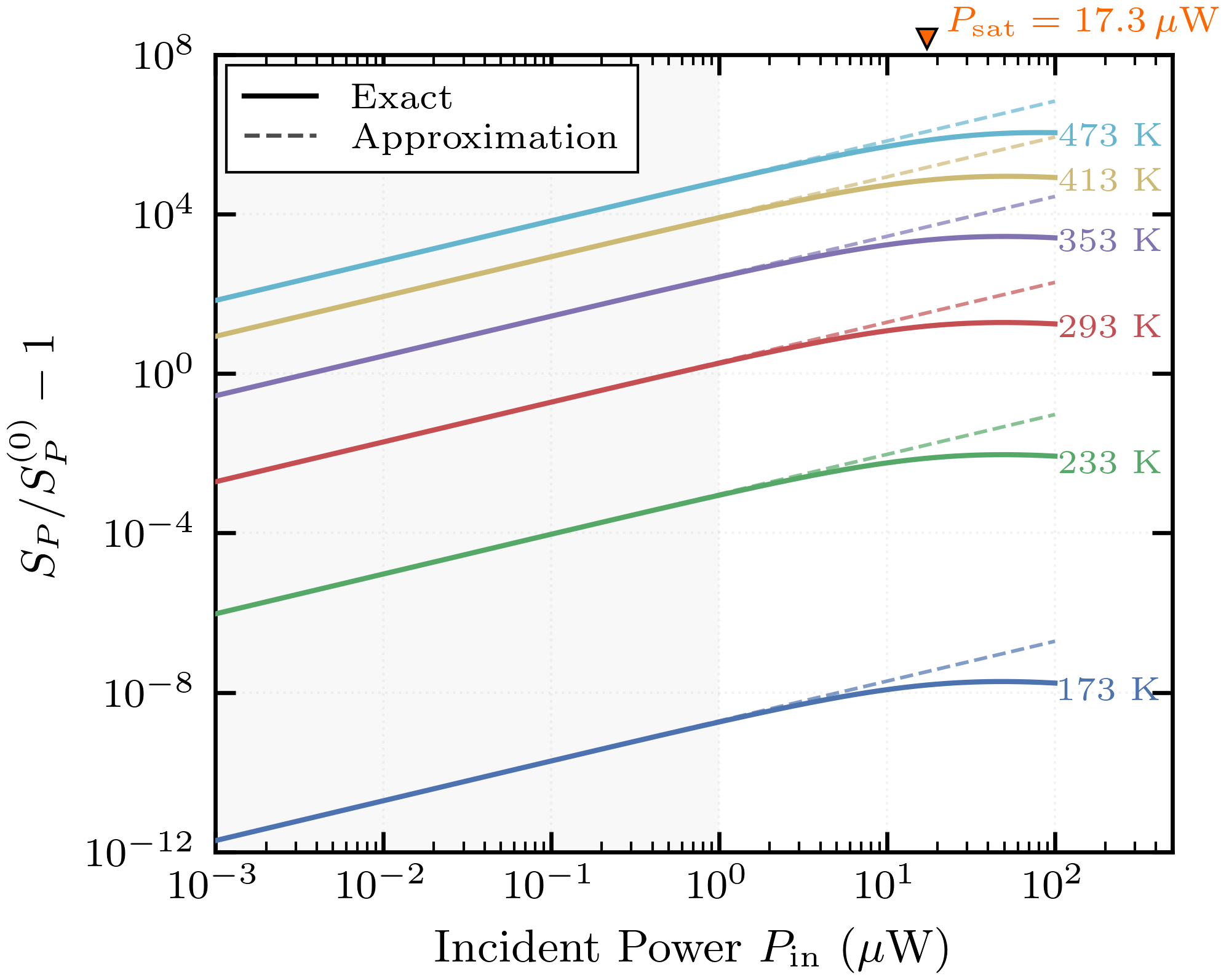}
\caption{PSD ratio $S_P/S_P^{(0)}-1$ as a function of incident power $P_{\mathrm{in}}$ for several atomic vapor temperatures (173--473\,K), computed using the cesium parameters from Table~\ref{tab:parameter}. The shaded region indicates the weak‑probe regime where the saturation parameter $x\ll1$ (see Appendix~\ref{app:master}); the room‑temperature saturation power $P_{\mathrm{sat}}\approx 17\,\mu\mathrm{W}$ is marked by the red arrow $\blacktriangledown$.}
\label{fig:scaling}
\end{figure}

Figure~\ref{fig:scaling} illustrates this linear scaling at several representative temperatures spanning 173--473\,K. For $P_{\mathrm{in}} \lesssim 1\,\mu\mathrm{W}$ (shaded region), i.e., the weak‑probe regime where the saturation parameter $x\ll1$ (see Appendix~\ref{app:master} for the explicit definition of $x$), the single‑atom polarizability $\alpha_I$ is independent of $P_{\mathrm{in}}$, and Eq.~\eqref{eq:scaling2K} predicts a linear dependence of the excess noise on $P_{\mathrm{in}}$. The unit slope observed for all curves on these log‑log axes directly confirms this linear proportionality and supports the extraction of $\mathcal{K}_{\mathrm{v}}$ from a simple linear fit. As $P_{\mathrm{in}}$ approaches $P_{\mathrm{sat}}$, the onset of saturation gradually bends the curves away from linearity, setting a natural upper bound for the measurement window.

\subsubsection{Cold atom cloud}

In a cold atomic cloud, the atoms are confined by external potentials, so the atom number $N_{\mathrm{at}}$ is fixed during the measurement. The cloud volume $V_{\mathrm{cl}}$ is determined from absorption imaging~\cite{reinaudi2007strong,vibel2024spatial,wolswijk2025trapping}. We assume the probe beam fully encompasses the cloud and that the cloud size is small compared to the Rayleigh range, so the probe intensity is approximately uniform over the interaction volume.

\begin{table}[t]
  \setlength{\tabcolsep}{5pt}
  \small
  \caption{Parameters for the $\mathrm{D}_2$ transition of ${}^{133}\mathrm{Cs}$ used throughout this work.
  For the cold-atom mode, values correspond to a typical magneto-optical trap~\cite{metcalf1999laser}.}
  \label{tab:parameter}
  \begin{tabular*}{\linewidth}{@{\extracolsep{\fill}} c c c}
    \toprule
    \textbf{Parameter} & \multicolumn{2}{c}{\textbf{Value}} \\
    \midrule
    $\gamma_2/2\pi$                & \multicolumn{2}{c}{$5.223\,\mathrm{MHz}$} \\
    $\mu_{21}$                     & \multicolumn{2}{c}{$2.5817\,e a_0$} \\
    $\lambda$                      & \multicolumn{2}{c}{$852\,\mathrm{nm}$} \\
    $w_0$                          & \multicolumn{2}{c}{$1.0\,\mathrm{mm}$} \\
    $\Delta/2\pi$                  & \multicolumn{2}{c}{$0\,\mathrm{MHz}$} \\
    $P_{\mathrm{in}}$               & \multicolumn{2}{c}{$10^{-3}$--$1\,\mathrm{\mu W}$} \\
    \midrule
    \textbf{Parameter}              & \textbf{Warm vapor} & \textbf{Cold atoms} \\
    \midrule
    $L$                            & $10.0\,\mathrm{mm}$ & $4.5\,\mathrm{mm}$ \\
    $n$                            & from $T$            & $1\times10^{10}\,\mathrm{cm}^{-3}$ \\
    $N_{\mathrm{at}}$              & from $T$            & $\approx1.4\times10^{8}$ \\
    $\Delta t$                     & $\tau_{\mathrm{tr}}$ & $10\,\mathrm{ms}$ \\
    $\sigma_{\mathrm{coll}}$       & $1\times10^{-17}\,\mathrm{m}^2$ & $0$ \\
    \bottomrule
	\noalign{\vspace{3pt}}
	\multicolumn{3}{l}{\footnotesize Note: parameters listed without a specific mode are common to both.}
  \end{tabular*}
\end{table}

For integration time $\Delta t$, the photon-to-atom ratio is $\mathcal{R} = P_{\mathrm{in}}\Delta t/(\hbar\omega N_{\mathrm{at}})$. In practice, $\Delta t \ll \tau_{\mathrm{cl}}$ (where $\tau_{\mathrm{cl}}$ is the cloud $1/e$ lifetime, typically several seconds~\cite{lemonde1995opto,li2012crossed,he2009single}), so a choice $\Delta t \sim 10\,\mathrm{ms}$ readily ensures $N_{\mathrm{at}}$ remains effectively constant. The derivation proceeds in complete analogy with the atomic vapor case, leading to the slope
\begin{equation}\label{eq:Kc}
\mathcal{K}_{\mathrm{c}} = \frac{\mathcal{I}\Delta t}{\hbar\omega N_{\mathrm{at}}}
= \frac{\omega}{\epsilon_0^2\hbar c^2 \pi^2 w_0^4}\,N_{\mathrm{at}}\Delta t\,\operatorname{var}_v[\alpha_I],
\end{equation}
where we have used $L_{\mathrm{eff}}$ from Eq.~\eqref{eq:Leff} and $\bar{n}=N_{\mathrm{at}}/V_{\mathrm{cl}}$. The dependence on the cloud volume $V_{\mathrm{cl}}$ cancels out, so that $\mathcal{K}_{\mathrm{c}}$ does not require precise knowledge of the cloud size.

\subsection{Temperature extraction}

According to Eqs.~\eqref{eq:Kv} and~\eqref{eq:Kc}, the temperature is therefore encoded predominantly in the variance $\operatorname{var}_v[\alpha_I]$, whose explicit form determines the scaling of $\mathcal{K}$ with $T$. As established in Sec.~\ref{sec:twolevel}, under resonant excitation and in the weak-probe limit this variance reduces to a function of the single dimensionless parameter $b_0$, with the exact expression given by Eq.~\eqref{eq:var_res}. We now examine the two asymptotic limits of $b_0$, which correspond to distinct physical regimes; the detailed expansions are carried out in Appendix~\ref{app:asymp}.

\subsubsection{The $b_0\ll1$ and $b_0\gg1$ limits}

\emph{The $b_0\ll1$ regime.}
For atomic vapors near room temperature, one finds $b_0 \approx 0.016$ using the parameters for a typical alkali $\mathrm{D}_2$ transition (Table~\ref{tab:parameter}), thereby confirming that $b_0 \ll 1$. In this regime, the Doppler width far exceeds the homogeneous linewidth. Expanding $\operatorname{erfc}$ for small argument (Appendix~\ref{app:asymp}) yields
\begin{equation}\label{eq:var_vapor}
\operatorname{var}_v[\alpha_I] \approx \frac{\sqrt{2\pi}\,C_0^{2}}{b_0}
= \frac{\sqrt{2\pi}\,|\bm{\epsilon}\cdot\bm{d}_{21}|^{4}}{2\hbar^{2}kv_{\mathrm{th}}(\gamma_2+2\Gamma)}, \quad b_0 \ll 1.
\end{equation}

\emph{The $b_0\gg1$ regime.}
At ultralow temperatures, the Doppler width pales beside the natural linewidth $\gamma_2$, pushing $b_0$ into the opposite regime $b_0 \gg 1$. Taking $T = 1\,\mathrm{mK}$ as representative with the parameters in Table~\ref{tab:parameter}, one obtains $v_{\mathrm{th}}\approx 0.27\,\mathrm{m/s}$ and $k v_{\mathrm{th}}/(2\pi)\approx 0.2\,\mathrm{MHz}$, giving $b_0 \approx \gamma_2/(2k v_{\mathrm{th}}) \approx 10$. Using the asymptotic expansion of $\operatorname{erfc}$ for large argument (Appendix~\ref{app:asymp}) then gives
\begin{equation}\label{eq:var_cold}
\operatorname{var}_v[\alpha_I] \approx \frac{8C_0^{2}}{b_0^{6}} 
= \frac{128\,|\bm{\epsilon}\cdot\bm{d}_{21}|^{4}k^{4}v_{\mathrm{th}}^{4}}{\hbar^{2}(\gamma_2+2\Gamma)^{6}}, \quad b_0 \gg 1.
\end{equation}

\begin{figure}[htbp]
\centering
\includegraphics{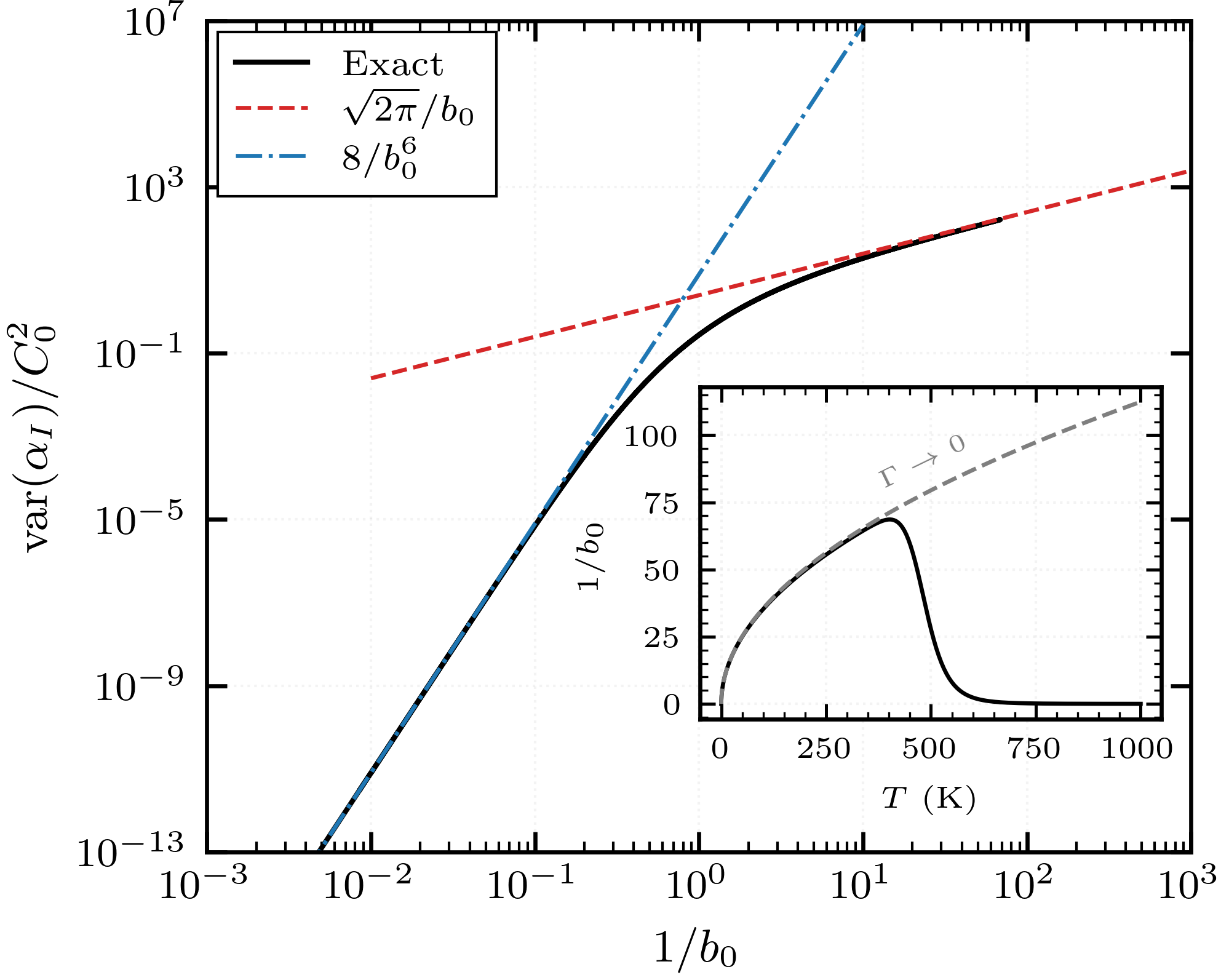}
\caption{Limiting behavior of the polarizability variance. Normalized variance $\operatorname{var}(\alpha_I)/C_0^2$ as a function of $1/b_0$ (or equivalently temperature) in log-log scale. Solid black curve: exact result from Eq.~\eqref{eq:var_res}; red dashed line: $b_0 \ll 1$ approximation $\propto 1/b_0$; blue dash-dotted line: $b_0 \gg 1$ asymptote $\propto 1/b_0^6$. The inset shows $1/b_0$ versus temperature $T$, with the black solid line including transit and collision broadening ($\Gamma_t,\Gamma_c$) and the gray dashed line corresponding to the ideal case $\Gamma\to0$ (i.e., $\Gamma_t=\Gamma_c=0$).}
\label{fig:variance}
\end{figure}

Figure~\ref{fig:variance} displays the exact variance alongside both limits. The normalized variance $\operatorname{var}(\alpha_I)/C_0^2$ is plotted against $1/b_0$, which scales directly as $\sqrt{T}$ when $\Gamma \ll \gamma_2$. The exact expression (solid black curve) transitions smoothly from the $1/b_0$ approximation (red dashed) for $b_0\ll1$ to the $1/b_0^6$ asymptote (blue dash-dotted) for $b_0\gg1$. 

For $b_0 \ll 1$, the Doppler width dominates; the atomic coherence is rapidly scrambled by Doppler shifts within the transit time $\tau_{\mathrm{tr}}$, and the variance approaches the $1/b_0$ scaling. For $b_0 \gg 1$, the natural linewidth $\gamma_2$ dominates the atomic response; the velocity distribution is effectively frozen, and the variance follows the $1/b_0^6$ decay. The inset compares the exact $1/b_0$--$T$ dependence (black solid curve, including full transit-time and collision broadening) with the idealized limit $\Gamma \to 0$ (gray dashed curve). Where $\Gamma \ll \gamma_2$, the two curves coincide and $b_0 \approx \gamma_2/(2k v_{\mathrm{th}})$. Above $\sim350$ K, collision broadening grows rapidly because the vapor density rises exponentially with $T$, driving $\Gamma_c \propto e^{-B/T}/\sqrt{T}$ upward and forcing the exact curve away from the $\Gamma \to 0$ limit---hence the breakdown of the $b_0\ll1$ approximation.

\subsubsection{Temperature scalings}

Inserting the approximate and asymptotic forms of the variance obtained above into the slope expressions~\eqref{eq:Kv} and~\eqref{eq:Kc} reveals how $\mathcal{K}$ depends on temperature in the two physical settings.

\emph{Atomic vapor.}
Substituting the $b_0\ll1$ approximation Eq.~\eqref{eq:var_vapor} into Eq.~\eqref{eq:Kv} yields
\begin{equation}
\mathcal{K}_{\mathrm{v}} = \frac{L}{w_0} \frac{|\bm{\epsilon}\cdot\bm{d}_{21}|^{4}}{2\epsilon_0^{2}\hbar^3 c} \frac{n(T)}{[\gamma_2+2\Gamma(T)]\,v_{\mathrm{th}}^2(T)}.
\end{equation}
Considering $n(T)=P_{\mathrm{v}}(T)/(k_B T)$ where $P_{\mathrm{v}}(T)$ is the saturated vapor pressure of the atomic species, and noting that $\Gamma \ll \gamma_2$, the slope scales as
\begin{equation}\label{eq:Kvsim}
\mathcal{K}_{\mathrm{v}} \propto \frac{P_{\mathrm{v}}(T)}{T^{2}}.
\end{equation}
Thus, $\mathcal{K}_{\mathrm{v}}$ inherits the exponential temperature dependence of the saturated vapor pressure, $\propto \exp(-B/T)$~\cite{steck2025Cesium}, which may provide high sensitivity in a limited temperature range, particularly near room temperature. In atomic vapors, the applicable temperature range is bounded at the low end by the vanishing vapor pressure (below $\sim150$ K for Cs) and at the high end by the breakdown of the $b_0 \ll 1$ approximation noted earlier. This range coincides with the operational window of DBT, yet GNT demands no pressure extrapolation.

\emph{Cold atom cloud.}
Substituting the $b_0\gg1$ asymptote Eq.~\eqref{eq:var_cold} into Eq.~\eqref{eq:Kc} gives
\begin{equation}
\mathcal{K}_{\mathrm{c}} = \frac{1}{(\pi w_0^2)^2}\frac{128\,|\bm{\epsilon}\cdot\bm{d}_{21}|^{4}\,\omega^5}{\epsilon_0^2\hbar^3 c^6}\frac{N_{\mathrm{at}}\Delta t\,v_{\mathrm{th}}^{4}}{(\gamma_2+2\Gamma)^{6}}.
\end{equation}
At millikelvin temperatures, $\Gamma \ll \gamma_2$, and with $v_{\mathrm{th}} \propto \sqrt{T}$ we obtain
\begin{equation}\label{eq:Kcsim}
\mathcal{K}_{\mathrm{c}} \propto N_{\mathrm{at}}\,\Delta t\,v_{\mathrm{th}}^4 \propto N_{\mathrm{at}}\,\Delta t\,T^2.
\end{equation}
The $T^2$ scaling implies that $\mathcal{K}_{\mathrm{c}}$ decreases rapidly as $T \to 0$. Nevertheless, both $N_{\mathrm{at}}$ and $\Delta t$ are experimentally tunable, so that a measurable signal can be maintained even at microkelvin temperatures. The slope $\mathcal{K}_{\mathrm{c}}$ is determined by $N_{\mathrm{at}}$, $\Delta t$, and the beam waist $w_0$, and is largely independent of the detailed density profile as long as the probe beam sufficiently overlaps the atomic cloud.

Taken together, Eqs.~\eqref{eq:Kvsim} and~\eqref{eq:Kcsim} show that the same underlying variance asymptotics produce qualitatively distinct temperature scalings in the two physical settings---exponential in the vapor regime, a power law in the cold-atom regime---both of which follow directly from the unified scaling law Eq.~\eqref{eq:scaling}.

\begin{figure*}[t]
\centering
\includegraphics{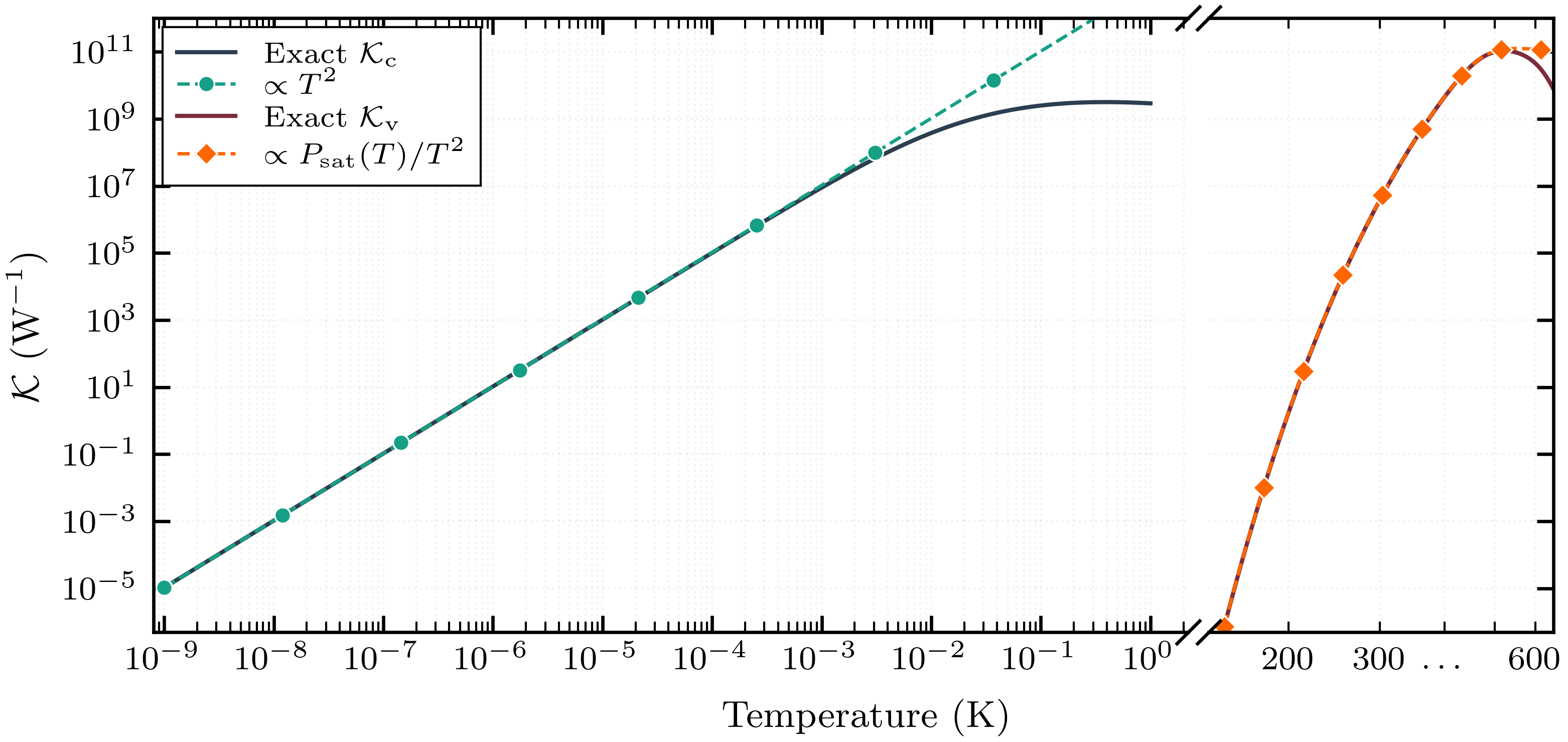}
\caption{Slope $\mathcal{K}$ as a function of temperature. Left: cold atoms (1\,nK--1\,K), with exact $\mathcal{K}_{\mathrm{c}}$ (dark solid) and $b_0\gg1$ asymptote $\propto T^{2}$ (dashed with circular markers). Right: atomic vapor (150--650\,K), with exact $\mathcal{K}_{\mathrm{v}}$ (dark solid) and $b_0\ll1$ approximation $\propto P_{\mathrm{v}}(T)/T^{2}$ (dashed with diamond markers). The temperature axis is broken between 1\,K and 150\,K. Both axes use log-log scale. Parameters are listed in Table~\ref{tab:parameter}.}
\label{fig:K}
\end{figure*}

Figure~\ref{fig:K} illustrates the thermometric performance of GNT in both implementations, using the parameters listed in Table~\ref{tab:parameter}. The left panel covers the cold-atom data, where $\mathcal{K}_{\mathrm{c}}$ adheres to the $T^{2}$ scaling down to nanokelvin temperatures---well into the domain where DBT is inoperative. Above $\sim 1\,\mathrm{mK}$, the $b_0\gg1$ asymptote (dashed with circular markers) departs from the exact curve (dark solid) as Doppler broadening becomes comparable to the natural linewidth; throughout the cold-atom regime, however, the $T^{2}$ scaling remains accurate. The right panel presents the atomic vapor implementation, where $\mathcal{K}_{\mathrm{v}}$ varies by over thirteen orders of magnitude across the $150$--$350\,\mathrm{K}$ range, reflecting the exponential sensitivity inherited from the saturated vapor pressure. Throughout this interval, the $b_0\ll1$ approximation (dashed with diamond markers) agrees with the exact result (dark solid) for room-temperature vapors. While the cold-atom curves assume parameters typical of a magneto-optical trap: $L_{\mathrm{eff}}=4.5\,\mathrm{mm}$, $n=1\times10^{10}\,\mathrm{cm}^{-3}$, $N_{\mathrm{at}}=1.4\times10^{8}$, and $\Delta t=10\,\mathrm{ms}$~\cite{metcalf1999laser,wolswijk2025trapping}. Under these conditions, $\mathcal{K}_{\mathrm{c}}$ varies by twelve orders of magnitude from $1\,\mathrm{nK}$ to $1\,\mathrm{mK}$. Taken together, the two segments confirm that the same underlying scaling relation applies to both thermal vapors and ultracold atomic clouds. Experimentally, $\mathcal{K}$ is obtained by sweeping $P_{\mathrm{in}}$ and fitting the PSD ratio. In the two limiting regimes, the temperature is determined from the scaling laws [Eqs.~\eqref{eq:Kvsim} and~\eqref{eq:Kcsim}]; outside these regimes, it is obtained from the exact theoretical curves.

\section{\label{sec:discussion} Discussion}

The unified scaling law derived in Sec.~\ref{sec:stat} and the analytical expressions for $\mathcal{K}$ obtained in Sec.~\ref{sec:temp} relate the measured slope $\mathcal{K}$ directly to thermodynamic temperature. Since $\mathcal{K}$ is extracted from a linear fit to noisy PSD data, the achievable temperature precision is governed by the statistics of $\mathcal{K}$. Within the Fisher information framework~\cite{degen2017quantum}, each spectral scan processes a block of $N$ time-domain samples acquired at intervals $\Delta t$ and produces a single estimate $Y = S_P/S_P^{(0)}$ of the PSD ratio, where $S_P^{(0)}$ is constant for given incident optical powers. For a white noise process, the single-scan periodogram $Y$ follows an exponential distribution with variance equal to the square of its mean~\cite{percival2020spectral,stoica2005spectral}; in the present context, this implies $\operatorname{var}[Y] = \mu_Y^{2}$, where $\mu_Y = \operatorname{E}[Y]$ denotes the true but unknown PSD ratio at a given $P_{\mathrm{in}}$.

Averaging over $M$ such independent scans yields the sample mean $\bar{Y} = M^{-1}\sum_{i=1}^{M} Y_i$, which is an unbiased estimator of $\mu_Y$ with variance $\operatorname{var}[\bar{Y}] = \operatorname{var}[Y]/M = \mu_Y^{2}/M$. Since $\mu_Y$ is unknown, the measured sample mean $\bar{Y}$ is used in its place, giving $\operatorname{var}[\bar{Y}] = \bar{Y}^{2}/M$. For sufficiently large $M$, the central limit theorem guarantees that $\bar{Y}$ is approximately Gaussian with mean $\mu_{\bar{Y}}(T) = 1 + P_{\mathrm{in}}\,\mathcal{K}(T)$ and variance $\bar{Y}^{2}/M$.

Repeating this procedure at $N_p$ distinct incident powers yields $N_p$ independent data points $\{\bar{Y}_j\}_{j=1}^{N_p}$. As illustrated in Fig.~\ref{fig:scaling}, the $N_p$ incident powers can be uniformly spaced from $P_{\mathrm{min}} \sim 10^{-3}\,\mu\mathrm{W}$ to $P_{\mathrm{max}} \sim 1\,\mu\mathrm{W}$. In this regime, the excess noise contributes only a small fraction, such that $\bar{Y}_j - 1 \sim O(1)$ and each data point has approximately the same variance $\bar{Y}_j^{2}/M \approx 1/M$. With $\bar{Y}_j \sim \mathcal{N}(\mu_{\bar{Y}_j}, \bar{Y}_j^{2}/M)$, the Fisher information for $N_p$ independent measurements~\cite{degen2017quantum} is
\begin{equation}
\mathcal{F}(T) = \sum_{j=1}^{N_p} \frac{M}{\bar{Y}_j^{2}} \left( \frac{\partial \mu_{\bar{Y}_j}}{\partial T} \right)^{2}
\sim N_p M \left( \frac{d\ln\mathcal{K}}{dT} \right)^2,
\end{equation}
where $\sum_j P_{\mathrm{in},j}^2 \approx (N_p/3) P_{\mathrm{max}}^2$ and $P_{\mathrm{max}} \mathcal{K} \sim O(1)$ have been utilized. Consequently, the Cram\'er--Rao inequality $\delta T \ge 1/\sqrt{\mathcal{F}(T)}$ bounds the temperature precision. For practical estimates, we take $N_p \sim 100$ and $M \sim 10^{4}$. For the vapor case, $\mathcal{K}_{\mathrm{v}} \propto P_{\mathrm{v}}(T)/T^{2} \propto e^{-B/T}/T^{2}$, giving $|d\ln\mathcal{K}_{\mathrm{v}}/dT| \approx B/T^{2}$ when $T \ll B$. The Cram\'er--Rao bound yields $\delta T \gtrsim T^{2}/(B\sqrt{N_p M})$. With $B \approx 9200\,\mathrm{K}$ for cesium~\cite{steck2025Cesium} and $T = 300\,\mathrm{K}$, we find $\delta T \sim 10\,\mathrm{mK}$. For the cold‑atom case, $\mathcal{K}_{\mathrm{c}} \propto T^{2}$ gives $|d\ln\mathcal{K}_{\mathrm{c}}/dT| = 2/T$, leading to $\delta T \gtrsim T/(2\sqrt{N_p M})$, which at $T = 10\,\mu\mathrm{K}$ evaluates to $\delta T \sim 5\,\mathrm{nK}$. Both estimates follow directly from the Cram\'er--Rao bound and illustrate the theoretical temperature precision achievable with realistic measurement parameters.

While the above estimates theoretically establish the statistical precision achievable, the experimental realization of GNT would impose additional requirements on the optical setup and the atomic ensemble. Both slope expressions involve the beam waist $w_0$, with a particularly steep $1/w_0^{4}$ scaling in the cold-atom case, and the vapor result additionally depends on the cell length $L$. These geometric quantities must therefore be measured with high precision. Further, technical noise sources such as laser intensity fluctuations and frequency noise must be systematically suppressed; balanced detection and modulation techniques can mitigate these effects, and a quantitative noise budget will be essential for experimental implementation. For cold atoms, the present treatment assumes a uniform probe beam fully encompassing the cloud; Gaussian density profiles and finite optical depths may require a more refined spatial treatment for high-precision metrology. Notably, quantum statistical effects are expected to modify the velocity distribution at sub-microkelvin temperatures, leading to deviations from the classical $T^{2}$ scaling. Extending GNT into this quantum domain would be a natural direction for future investigation. Conceivably, the noise spectrum analyzed here carries information beyond temperature: time- and frequency-resolved measurements of the granularity signal may provide access to relaxation dynamics, critical fluctuations near phase transitions, and interaction-induced correlations in the atomic ensemble.

\section{\label{sec:conclusion} Conclusion}

We have introduced granularity noise thermometry (GNT), an optical thermometry scheme that extracts temperature from the power spectral density of light transmitted through an atomic ensemble. A single scaling law relates the excess noise to the photon-to-atom ratio in both thermal vapors and ultracold atomic clouds. In the vapor regime, the slope scales as $P_{\mathrm{v}}(T)/T^{2}$; in the cold‑atom regime, it scales as $T^{2}$ down to microkelvin temperatures, below which quantum statistical effects are expected to modify the velocity distribution. The method uses a laser, a photodetector, and a spectrum analyzer, provided technical noise can be suppressed effectively. By exploiting atom-number granularity rather than motional line broadening, GNT provides a noise-based approach to thermometry. The present framework may be extended to other platforms where discrete scatterers interact with a probe field, and to the quantum regime where the classical Maxwell–Boltzmann distribution must be replaced by the quantum statistics.

\begin{acknowledgments}
This work was supported by the National Natural Science Foundation of China (Grants No.~12304545 and No.~12475042) and the National Key Research and Development Program of China (Grant No.~2023YFF0718400). The authors acknowledge the use of artificial-intelligence-powered language editing tools to improve the readability and grammatical accuracy of this article while preserving scientific rigor and technical terminology.
\end{acknowledgments}

\appendix

\section{\label{app:psd} Power spectral density ratio}

This appendix derives the scaling law linking the power spectral density (PSD) of the transmitted light to the atomic granularity noise. The notation follows Sec.~\ref{sec:stat}: the transmitted power is $P_{\mathrm{tr}} = P_{\mathrm{in}} e^{-a\bar{\chi}_I}$ with $a = 2\pi L/\lambda$, $\bar{\chi}_I$ is the sample-averaged susceptibility, $\bar{N}_{\mathrm{ph}} = P_{\mathrm{in}}\Delta t/(\hbar\omega)$ is the mean detected photon number per interval $\Delta t$, and $\sigma_I^2 = \mathcal{V}_I/\bar{N}_{\mathrm{at}}$ is the variance of $\bar{\chi}_I$ with $\mathcal{V}_I = (n^2/\epsilon_0^2)\operatorname{var}_v[\alpha_I]$.

The photon count $\xi$ and the susceptibility $\bar{\chi}_I$ are statistically independent. For a coherent-state probe, $\xi$ is Poisson distributed with $\operatorname{E}[\xi]=\operatorname{var}[\xi]=\bar{N}_{\mathrm{ph}}$. For large $\bar{N}_{\mathrm{at}}$, $\bar{\chi}_I$ is approximately Gaussian with mean $\mu_I$ and variance $\sigma_I^2 = \mathcal{V}_I/\bar{N}_{\mathrm{at}}$.

Define the random variable $\zeta = \xi \exp(-a\bar{\chi}_I)$. Using the law of total variance and the independence of $\xi$ and $\bar{\chi}_I$, we obtain
\begin{align}
\operatorname{E}[\zeta] &= \bar{N}_{\mathrm{ph}} \operatorname{E}[e^{-a\bar{\chi}_I}], \\
\operatorname{var}[\zeta] &= \bar{N}_{\mathrm{ph}} \operatorname{E}[e^{-a\bar{\chi}_I}] + \bar{N}_{\mathrm{ph}}^2 \operatorname{var}[e^{-a\bar{\chi}_I}].
\end{align}
For a Gaussian $\bar{\chi}_I$, the required exponential moments are
\begin{align}
\operatorname{E}[e^{-a\bar{\chi}_I}] &= e^{-a\mu_I + \frac12 a^2\sigma_I^2}, \\
\operatorname{E}[e^{-2a\bar{\chi}_I}] &= e^{-2a\mu_I + 2a^2\sigma_I^2}.
\end{align}
Substituting these and collecting terms gives
\begin{equation}
\operatorname{var}[\zeta] = e^{-2a\mu_I + a^2\sigma_I^2} \bar{N}_{\mathrm{ph}} \Bigl[ (1 + \bar{N}_{\mathrm{ph}}) e^{a^2\sigma_I^2} - \bar{N}_{\mathrm{ph}} \Bigr].
\end{equation}
The measured power in a single interval is $P = (\hbar\omega/\Delta t)\zeta$. For a white noise process sampled at intervals $\Delta t$, the single-sided PSD is $S_P = 2\Delta t \operatorname{var}[P] = 2(\hbar\omega)^2/\Delta t \operatorname{var}[\zeta]$. The shot-noise limit $S_P^{(0)}$ is obtained by setting $\sigma_I \to 0$, yielding $S_P^{(0)} = 2\hbar\omega P_{\mathrm{in}} e^{-2a\mu_I}$. The exact PSD ratio is therefore
\begin{equation}
\frac{S_P}{S_P^{(0)}} = e^{a^2\sigma_I^2} \Bigl[ (1 + \bar{N}_{\mathrm{ph}}) e^{a^2\sigma_I^2} - \bar{N}_{\mathrm{ph}} \Bigr].
\end{equation}
For $\bar{N}_{\mathrm{at}} \gg 1$, we have $\sigma_I^2 \propto 1/\bar{N}_{\mathrm{at}} \ll 1$ and hence $a^2\sigma_I^2 \ll 1$. Expanding to first order in $\sigma_I^2$,
\begin{equation}
\frac{S_P}{S_P^{(0)}} \approx 1 + \bar{N}_{\mathrm{ph}}\, a^2\sigma_I^2.
\end{equation}
Substituting $\mathcal{R} = \bar{N}_{\mathrm{ph}}/\bar{N}_{\mathrm{at}}$ and $\mathcal{I} = a^2 (n^2/\epsilon_0^2)\operatorname{var}_v[\alpha_I] = a^2 \bar{N}_{\mathrm{at}}\sigma_I^2$ yields the scaling law $S_P/S_P^{(0)} = 1 + \mathcal{R}\mathcal{I}$, which is Eq.~\eqref{eq:scaling} in the main text.

\section{\label{app:master} Two-level atom: master equation, polarizability, and statistical moments}

This appendix provides the complete derivation of the analytical single-atom polarizability for a two-level atom and its statistical moments, supporting the results presented in Sec.~\ref{sec:twolevel}.

\subsection{Master equation and steady-state solution}

We start from the master equation including spontaneous decay, transit-time broadening,
and collision-induced dephasing~\cite{sagle1996measurement,jabbour1995measurement,pitz2010pressure,steck2025Cesium}:
\begin{equation}
\frac{\partial \rho}{\partial t} = \frac{1}{i\hbar}[H_R,\rho] + \mathcal{L}_{\mathrm{decay}}\rho
+ \mathcal{L}_{\mathrm{tran}}\rho + \mathcal{L}_{\mathrm{coll}}\rho,
\end{equation}
with
\begin{align}
\mathcal{L}_{\mathrm{decay}}\rho &= \gamma_2\left(\sigma_{12}\rho\sigma_{12}^{\dagger}
- \frac12\{\sigma_{12}^{\dagger}\sigma_{12},\rho\}\right),\\
\mathcal{L}_{\mathrm{tran}}\rho &= -\Gamma_t(\rho_{12}|1\rangle\langle2|+\rho_{21}|2\rangle\langle1|),\\
\mathcal{L}_{\mathrm{coll}}\rho &= -\Gamma_{c}(\rho_{12}|1\rangle\langle2|+\rho_{21}|2\rangle\langle1|),
\end{align}
and $\sigma_{12}=|1\rangle\langle2|$. The three rates $\gamma_2$, $\Gamma_t$, and $\Gamma_c$
are defined in Sec.~\ref{sec:twolevel}, with $\Gamma = \Gamma_t + \Gamma_c$ the total homogeneous dephasing rate. To solve for the steady state, we vectorize the density matrix. In the basis
$\{|1\rangle,|2\rangle\}$, the density matrix is
\begin{equation}
\rho = \begin{pmatrix}\rho_{11}&\rho_{12}\\\rho_{21}&\rho_{22}\end{pmatrix}.
\end{equation}
Vectorizing $\rho$ as $|\rho\rangle = (\rho_{11},\rho_{12},\rho_{21},\rho_{22})^{\mathsf{T}}$,
the Liouvillian $G$ satisfies $\partial_t|\rho\rangle=G|\rho\rangle$. Explicitly,
\begin{widetext}
\begin{equation}\label{eq:Gmatrix}
G = \begin{pmatrix}
0 & \frac{i\Omega}{2} & -\frac{i\Omega}{2} & \gamma_2 \\
\frac{i\Omega}{2} & -i(\Delta-kv) - \frac{\gamma_2}{2} - \Gamma_t - \Gamma_{c} & 0 & -\frac{i\Omega}{2} \\
-\frac{i\Omega}{2} & 0 & i(\Delta-kv) - \frac{\gamma_2}{2} - \Gamma_t - \Gamma_{c} & \frac{i\Omega}{2} \\
0 & -\frac{i\Omega}{2} & \frac{i\Omega}{2} & -\gamma_2
\end{pmatrix}.
\end{equation}
\end{widetext}
Decomposing $G=G_0+vG_v$, where $G_0$ contains the terms independent of $v$ and $G_v$
contains the $v$-dependent diagonal entries, the steady-state solution is obtained via the
Drazin inverse~\cite{miller2024rydiqule,nagib2025fast}:
\begin{equation}
|\rho_v\rangle = \left[I + vG_0^{-1}G_v\right]^{-1}|\rho_0\rangle,
\end{equation}
where $|\rho_0\rangle$ is the zero-velocity steady-state solution. The Drazin inverse $G_0^{-1}$
is constructed by projecting out the null space:
\begin{equation}
G_0^{-1} = \left(G_0 + |\rho_0\rangle\langle l_0|\right)^{-1} - |\rho_0\rangle\langle l_0|,
\end{equation}
with $\langle l_0|$ the left eigenvector satisfying $\langle l_0|\rho_0\rangle = 1$.
Diagonalizing $M = v_{\mathrm{th}} G_0^{-1} G_v$ yields a representation in terms of simple
poles:
\begin{equation}
\rho_{21}(v) = \sum_{k=1}^{4} \frac{c_k}{1 + \lambda_k \xi},
\end{equation}
which for the two-level system truncates to two terms, giving the explicit rational
expression Eq.~\eqref{eq:rho21v} in the main text.

\subsection{Partial-fraction decomposition}

To obtain a form suitable for integration over the velocity distribution, we perform a
partial-fraction decomposition of $\rho_{21}(v)$. Introducing $v_{\mathrm{th}}$ and the
dimensionless velocity $\xi = v/v_{\mathrm{th}}$, Eq.~\eqref{eq:rho21v} becomes a rational
function of $\xi$:
\begin{equation}
\rho_{21}(\xi)=\frac{M_1\xi+M_0}{D_2\xi^{2}+D_1\xi+D_0},
\end{equation}
with
\begin{align*}
M_1 &= -2\gamma_2\Omega k v_{\mathrm{th}},\\
M_0 &= \gamma_2\Omega[2\Delta - i(\gamma_2+2\Gamma)],\\
D_2 &= 4\gamma_2 k^{2} v_{\mathrm{th}}^{2},\\
D_1 &= -8\gamma_2 k v_{\mathrm{th}}\Delta,\\
D_0 &= (\gamma_2+2\Gamma)(\gamma_2^{2}+2\gamma_2\Gamma+2\Omega^{2}) + 4\gamma_2\Delta^{2},
\end{align*}
with $\Gamma = \Gamma_t + \Gamma_c$ as defined above. The denominator roots $r_{1,2}$ are
given by Eq.~\eqref{eq:r12} of the main text. Solving for the residues $A,B$ via
$A+B = M_1/D_2$ and $A r_2 + B r_1 = -M_0/D_2$ yields
\begin{align}
A &= -\frac{\Omega}{4k v_{\mathrm{th}}}\left(1+\frac{1}{\sqrt{1+x}}\right),\\
B &= -\frac{\Omega}{4k v_{\mathrm{th}}}\left(1-\frac{1}{\sqrt{1+x}}\right),
\end{align}
where $x$ is the saturation parameter [Eq.~\eqref{eq:x}]. The coherence thus takes the
pole form
\begin{equation}\label{eq:rho21_app}
\rho_{21}(\xi) = \frac{A}{\xi-r_1}+\frac{B}{\xi-r_2}.
\end{equation}

\subsection{Polarizability from the steady-state coherence}

The single-atom polarizability follows from the linear response of the induced dipole moment
to the applied electric field. For a two-level atom without permanent dipole moment, the
induced dipole moment in the steady state is given by
\begin{equation}
\langle \bm{d} \rangle = \mathrm{Tr}(\rho \bm{d}) = \bm{d}_{12}\rho_{12} + \bm{d}_{21}\rho_{21},
\end{equation}
where $\bm{d}_{ij} = \langle i | \bm{d} | j \rangle$ are the dipole matrix elements. The probe
field is taken as
\begin{equation}
\bm{E}(t) = \bm{\epsilon} \mathcal{E}^{+} e^{-i\omega t} + \text{c.c.},
\end{equation}
with $\mathcal{E}^{+}$ the positive-frequency amplitude. Equating the two expressions and
solving for $\alpha$ yields
\begin{equation}
\alpha(v) = \frac{2|\bm{\epsilon}\cdot\bm{d}_{21}|^{2}}{\hbar\Omega}\,\rho_{21}(v),
\end{equation}
where we have used $\Omega = 2\langle 2|\bm{\epsilon}\cdot\bm{d}|1\rangle \mathcal{E}^{+}/\hbar$
and the fact that $\rho_{21}$ is proportional to $\Omega$, ensuring that $\alpha$ is independent
of the Rabi frequency in the linear regime. Substituting the partial-fraction form of
$\rho_{21}(\xi)$  Eq.~\eqref{eq:rho21_app} into the above expression, we obtain the compact
presentation Eq.~\eqref{eq:alpha} with the coefficients $c_{1,2}$ given in Eq.~\eqref{eq:c12}
of the main text.

\subsection{\label{subapp:moments} Statistical moments}

The statistical moments of $\alpha(\xi)$ over the Maxwell-Boltzmann distribution,
required for evaluating $\operatorname{var}_v[\alpha_I]$ in Sec.~\ref{sec:twolevel},
reduce to integrals of rational functions. These are expressed compactly via the plasma
dispersion function $Z(z)$~\cite{fried1961plasma}, defined as
\[
Z(z) = \frac{1}{\sqrt{\pi}}\int_{-\infty}^{\infty} \frac{e^{-t^{2}}}{t-z}dt,\quad \Im z>0,
\]
and analytically continued to the whole complex plane. For the Maxwell-Boltzmann
distribution $\phi(\xi)=e^{-\xi^{2}/2}/\sqrt{2\pi}$, the fundamental integrals are
\begin{align}
\int_{-\infty}^{\infty}\frac{1}{\xi-r}\phi(\xi)d\xi &= \frac{1}{\sqrt{2}}Z\!\left(\frac{r}{\sqrt{2}}\right),\\
\int_{-\infty}^{\infty}\frac{1}{(\xi-r)^{2}}\phi(\xi)d\xi &= -\frac{r}{\sqrt{2}}Z\!\left(\frac{r}{\sqrt{2}}\right) - 1,\\
\int_{-\infty}^{\infty}\frac{1}{(\xi-r_1)(\xi-r_2)}\phi(\xi)d\xi &= \frac{1}{\sqrt{2}(r_1-r_2)}\left(Z_1-Z_2\right),
\end{align}
where $Z_k=Z(r_k/\sqrt{2})$. The second line follows from $Z'(z)=-2[1+z Z(z)]$.

With $\alpha(\xi)$ expressed as Eq.~\eqref{eq:alpha}, the squared modulus and square of the
polarizability decompose as
\begin{align*}
|\alpha|^2 &= \frac{c_1^{2}+c_2^{2}}{(\xi-r_1)(\xi-r_2)} + c_1c_2\!\left[\frac{1}{(\xi-r_1)^{2}}+\frac{1}{(\xi-r_2)^{2}}\right],\\
\alpha^{2} &= \frac{c_1^{2}}{(\xi-r_1)^{2}} + \frac{c_2^{2}}{(\xi-r_2)^{2}} + \frac{2c_1c_2}{(\xi-r_1)(\xi-r_2)}.
\end{align*}
Taking expectations using the above integrals yields the closed-form moments
\begin{align}
\mathbb{E}[\alpha] &= \frac{c_1}{\sqrt{2}}Z_1 + \frac{c_2}{\sqrt{2}}Z_2, \label{eq:Ealpha}\\
\mathbb{E}[|\alpha|^2] &= \frac{c_1^2 + c_2^2}{\sqrt{2}\,(r_1-r_2)}\left(Z_1 - Z_2\right) \nonumber\\
&\quad - c_1c_2\!\left(2 + \frac{1}{\sqrt{2}}(r_1 Z_1 + r_2 Z_2)\right), \label{eq:Eabs2}\\
\mathbb{E}[\alpha^2] &= \frac{2c_1c_2}{\sqrt{2}\,(r_1-r_2)}\left(Z_1 - Z_2\right) \nonumber\\
&\quad - (c_1^2 + c_2^2) - \frac{1}{\sqrt{2}}(c_1^2 r_1 Z_1 + c_2^2 r_2 Z_2). \label{eq:Ealpha2}
\end{align}
The variance of the imaginary part $\alpha_I$ follows as
\begin{equation}\label{eq:varI}
\operatorname{var}_v[\alpha_I] = \frac12\Bigl(\mathbb{E}[|\alpha|^{2}]-\Re\mathbb{E}[\alpha^{2}]\Bigr)-\left(\Im\mathbb{E}[\alpha]\right)^{2}.
\end{equation}

\section{\label{app:asymp} Asymptotic expansions of the polarizability variance}

The exact variance $\operatorname{var}_v[\alpha_I]$ under resonant, weak-probe conditions
is given by Eq.~\eqref{eq:var_res} of Sec.~\ref{sec:twolevel}. We now derive its leading
asymptotic behavior in the two limits relevant to the temperature extraction in
Sec.~\ref{sec:temp}: the high-temperature regime where Doppler broadening dominates
($b_0 \ll 1$), and the ultra-low-temperature regime where homogeneous broadening prevails
($b_0 \gg 1$). These limiting forms are quoted in the main text as
Eqs.~\eqref{eq:var_vapor} and~\eqref{eq:var_cold}.

\subsection{The $b_0 \ll 1$ Taylor expansion}

In the small-argument limit, the complementary error function admits the Taylor expansion
\begin{equation}
\operatorname{erfc}(y)=1-\frac{2y}{\sqrt{\pi}}\left(1-\frac{y^{2}}{3}+\frac{y^{4}}{10}-\frac{y^{6}}{42}+\cdots\right),\quad y\ll1.
\end{equation}
Substituting $y = b_0/\sqrt{2}$ and expanding the exponential prefactor $e^{b_0^2/2} = 1 + b_0^2/2 + b_0^4/8 + \cdots$, we obtain
\begin{align}
\Im Z_0 &= \sqrt{\pi}\,e^{\frac{b_0^{2}}{2}}\operatorname{erfc}\left(\frac{b_0}{\sqrt{2}}\right)\nonumber\\
&= \sqrt{\pi}\left(1+\frac{b_0^2}{2}+\frac{b_0^4}{8}+\cdots\right)\left(1-\frac{2b_0}{\sqrt{2\pi}}+\frac{b_0^{3}}{3\sqrt{2\pi}}+\cdots\right) \nonumber\\
&\approx \sqrt{\pi} - \sqrt{2}b_0 + \frac{\sqrt{\pi}}2b_0^2 - \frac{\sqrt{2}}{3}b_0^{3} + O(b_0^4).
\end{align}
Inserting this expansion into the exact variance formula Eq.~\eqref{eq:var_res} and retaining only the dominant term in the limit $b_0 \to 0$, the combination inside the brackets reduces to $\sqrt{\pi/2}/b_0$, yielding the simple asymptotic form
\begin{equation}
\operatorname{var}_v[\alpha_I] \approx 2C_0^{2}\cdot\frac{\sqrt{\pi/2}}{b_0} = \frac{\sqrt{2\pi}\,C_0^{2}}{b_0}.
\end{equation}
This $1/b_0$ scaling reflects the fact that the variance is controlled by the fraction of atoms whose Doppler shift lies within the homogeneous linewidth.

\subsection{The $b_0 \gg 1$ asymptotic expansion}

For large $b_0$, the complementary error function is more conveniently handled via its asymptotic expansion
\begin{equation}
\operatorname{erfc}(y) = \frac{e^{-y^{2}}}{y}\frac1{\sqrt{\pi}}\left(1 - \frac{1}{2y^{2}} + \frac{3}{4y^{4}} - \frac{15}{8y^{6}} + \cdots\right),\quad y\gg1.
\end{equation}
Setting $y = b_0/\sqrt{2}$ gives directly
\begin{equation}
\Im Z_0 = \frac{\sqrt{2}}{b_0}\left(1 - \frac{1}{b_0^{2}} + \frac{3}{b_0^{4}} - \frac{15}{b_0^{6}} + \cdots\right).
\end{equation}
To evaluate the variance, we need the combination $\frac{\Im Z_0}{\sqrt{2}}\left(\frac1{b_0}-b_0\right) + 1 - (\Im Z_0)^2$ appearing in Eq.~\eqref{eq:var_res}. Expanding each contribution in powers of $1/b_0$, we find
\begin{align}
\frac{\Im Z_0}{\sqrt{2}}\left(\frac1{b_0}-b_0\right) &= 
\left(\frac{1}{b_0} - \frac{1}{b_0^{3}} + \frac{3}{b_0^{5}} - \frac{15}{b_0^{7}} + \cdots\right)\left(-b_0 + \frac{1}{b_0}\right) \nonumber\\
&\approx -1 + \frac{2}{b_0^{2}} - \frac{4}{b_0^{4}} + \frac{18}{b_0^{6}} + O(b_0^{-7}),
\end{align}
and
\begin{align}
(\Im Z_0)^2 &= \frac{2}{b_0^{2}}\left(1 - \frac{2}{b_0^{2}} + \frac{7}{b_0^{4}} - \frac{36}{b_0^{6}} + \cdots\right)\nonumber\\
&\approx \frac{2}{b_0^{2}} - \frac{4}{b_0^{4}} + \frac{14}{b_0^{6}} + O(b_0^{-7}),
\end{align}
so that
\begin{equation}
1 - (\Im Z_0)^2 \approx 1 - \frac{2}{b_0^{2}} + \frac{4}{b_0^{4}} - \frac{14}{b_0^{6}} + O(b_0^{-7}).
\end{equation}
Adding the two contributions, the constant terms cancel exactly, as do the terms of order $b_0^{-2}$ and $b_0^{-4}$. The leading non-vanishing term is of order $b_0^{-6}$, and we obtain
\begin{equation}
\frac{\Im Z_0}{\sqrt{2}}\left(\frac1{b_0}-b_0\right) + 1 - (\Im Z_0)^2 \approx \frac{4}{b_0^{6}} + O(b_0^{-7}).
\end{equation}
Substituting this result back into Eq.~\eqref{eq:var_res} gives the asymptotic variance
\begin{equation}
\operatorname{var}_v[\alpha_I] \approx 2C_0^{2}\cdot\frac{4}{b_0^{6}} = \frac{8C_0^{2}}{b_0^{6}}.
\end{equation}
This steep $1/b_0^6$ decay arises because at ultralow temperatures the velocity distribution is effectively frozen on the scale of the homogeneous linewidth, leaving only a residual sensitivity to the tiny thermal spread. These expansions are employed in Sec.~\ref{sec:temp} to extract the temperature dependence of the slope $\mathcal{K}$ in both regimes.

\bibliography{GraThermoRef}

\end{document}